\documentclass[journal=jacsat,manuscript=article]{achemso}

\usepackage[version=3]{mhchem} 
\usepackage[separate-uncertainty=true]{siunitx} 
\usepackage{hyperref}
\usepackage{cleveref}

\author{Paul H. Bittorf}
\author{Filip Majstorovic}
\affiliation[Kiel University]
{Institute for Experimental and Applied Physics, Kiel University, 24118 Kiel, Germany}
\author{Pavel Ruchka}
\author{Harald Giessen}
\affiliation[Stuttgart University]
{4th Physics Institute and Research Center SCoPE, University of Stuttgart, 70569 Stuttgart, Germany}
\author{Nahid Talebi}
\affiliation[Kiel University]
{Institute for Experimental and Applied Physics, Kiel University, 24118 Kiel, Germany}
\alsoaffiliation[KiNSIS]
{Kiel Nano, Surface and Interface Science KiNSIS, Kiel University, 24118 Kiel, Germany}
\email{bittorf@physik.uni-kiel.de, talebi@physik.uni-kiel.de}

\title[Fiber CL Detector]
  {A Multi-Dimensional Cathodoluminescence Detector with 3D Printed Micro-Optics on a Fiber}

\begin{document}
\clearpage

\begin{abstract}
Cathodoluminescence, i.e. the radiation caused by the interaction of high-energy electron beams with matter, has gained a major interest in the analysis of minerals, semiconductors, and plasmonic resonances in nanoparticles. This radiation can either be coherent or incoherent, depending on the underlying interaction mechanism of electrons with nanostructured matter. Thanks to their high spatial resolution and large spectral excitation bandwidth, the evanescent near-field of a moving electron in a scanning electron microscope is used to probe locally photonic modes at the nanoscale, e.g., exciton or plasmon polaritons. The properties of these excitations can be analyzed through both spectral and temporal statistics of the emitted light. Here, we report on the technical design and implementation of a novel fiber-based cathodoluminescence detector for a scanning electron microscope. Moreover, we present first characterization measurements to prove the ability for raster scanning the cathodoluminescence emission using optical fibers with 3D printed micro-optics. The functionality and flexibility of this fiber-based detector is highlighted by resolving the spatial far-field distribution of the excited light, as well as cathodoluminescence spectroscopy and time-correlated single photon counting. Our findings pave the way for a better understanding of the characteristic of the light emitted from electron beams interacting with nanostructures and two-dimensional materials.
\end{abstract}
\clearpage

\section{Introduction}
\label{sec:introduction}

Electron microscopy and spectroscopy have merged to a versatile measurement technique for the investigation of excitons and bandgap excitations in semiconductors \cite{Yankovich2019,Vu2023,Vu2022,Davoodi2022,Taleb2023}. Due to the high spatial resolution of the electron microscopes and the large spectral bandwidth of the near-field of the moving electron interacting with matter, such material excitations are generally explorable at the nanoscale and within a broad spectral range, where the latter is mainly defined by the bandwidth of the analyzing path rather than the excitation mechanism itself. Depending on the underlying material excitations, light is emitted by the sample when electrons impact and interact with matter \cite{Abajo2021,Morimoto2018,Ryabov2016,Yacobi1986,Shiloh2022,Polman2019}. This so-called cathodoluminescence~(CL) can either be mutually incoherent or coherent with respect to the evanescent field of the incoming electron \cite{Coenen2017,Brenny2014,Feist2022}.

The spatio-temporal coherence properties of the radiation from electron beams interacting with matter are linked to the type of material excitations. In the case of metals, the accompanying evanescent field of the moving electrons polarizes the metal and induces an image charge inside, which forms a time-varying dipole while the electron is approaching \cite{Abajo2010}. After the electron passes the interface and the dipole is annihilated, coherent CL in the form of transition radiation is emitted, which exhibits a dipolar-like radiation profile in the far-field. Furthermore, electrons can couple to coherent collective modes, e.g., exciting propagating surface plasmon polaritons (SPP)\cite{Abajo2021,Talebi2017}. SPP generally do not contribute to the radiation, otherwise they interact with defects or other types of scatterers, where the distribution of such scatterers can be controlled for realizing coherent electron-driven photon sources \cite{Talebi2019,Christopher2020}. Generally, the primary electron enters the material and losses energy through a cascade of inelastic scattering interactions. Particularly in semiconductors, the secondary and backscattered electrons result in the excitation of bulk plasmons, phonons, electron-hole pairs in defect centers and excitons \cite{Toermae2015,Liu2016,Sang2021,Lummen2016,Garcia2020}. These excitations decay over time or recombine by emitting a photon, resulting in the emission of incoherent CL with a diffuse Lambertian radiation profile \cite{Brenny2014}. Nevertheless, strong coupling between excitons and cavity photons, for example in the case of self-hybridized exciton polaritons, exhibit coherent radiation properties as well, as shown recently via a sequential CL spectral interferometry technique \cite{Taleb2023,Taleb2022}.

For gathering emitted photons, inside transmission or scanning electron microscopes a parabolic mirror is usually installed near to the sample to collect the generated CL radiation and redirect it out of the electron microscope to a spectrometer or CCD camera for spectral and angular analysis \cite{Coenen2011,Kociak2014,Kociak2017}. The parabolic mirror features a solid acceptance angle for gathering the luminescence with a large collection efficiency. It is worth mentioning, that a few experiments already combine the mirror-based collection with optical fibers \cite{Tanabe2002,Meuret2021,Kim2021}, by focusing the gathered light into the fiber for further analysis, e.g., the Vulcan system from Gatan as a commercial application. However, the emitted luminescence is generally collected entirely by collimating the entire radiation, the exact places in the samples where the CL radiation are originated from remain unknown. It has been noted by Matsukata et al. \cite{Matsukata2022}, that this limitation can be circumvented by directly acquiring the emission image versus the electron beam position, where the resolution in determining the origin of the emission is given by the resolving power of the emission collection system.

Here, a novel fiber-based CL detector is developed by us to enable more selectivity in analyzing the origin of the emission from the sample. To further increase the numerical aperture, and thus the collection efficiency of the fiber, we 3D printed a lens onto its front side. Since the resolving power is related to the lens aperture size, exploiting a microlens with a significantly reduced aperture size enhances the collection efficiency. Compared to free-space optics, fiber-optic collection in CL offers improved mechanical stability, compactness, and alignment simplicity by avoiding free-space optics connected to the chamber of the microscope, which is especially beneficial in the constrained environment of electron microscopes.

Here, we provide the first proof-of-concept experiments, that demonstrate the efficiency of our fiber-based detector for performing spectroscopy and examining the radiation profile of the emission from sample, but also acquiring the photon statistics. This method provides us with multiple degrees of freedoms in analyzing the CL emission; namely, the scanning position of the electron beam ($x_\text{e}$, $y_\text{e}$), the scanning position of the fiber ($x_\text{f}$, $y_\text{f}$, $z_\text{f}$), the spectral distribution, as well as temporal statistics, where the latter is enabled by using the Hanbury-Brown and Twiss~(HBT) effect. All together the system provides a multidimensional data cube to analyze several aspects of the CL signal. To demonstrate the modularity and robustness of our fiber-based detection system, we extended our measurements beyond spectral imaging to include both the CL intensity dependence on electron beam current and second-order photon correlation~$g^{(2)}$ statistics. These measurements confirm that our system can reliably capture photon statistics in line with expected behaviors, such as photon bunching and current-dependent emission dynamics, as reported in prior studies. This versatility underscores the system’s suitability for a wide range of advanced CL experiments, including those involving quantum and time-resolved photonic phenomena. Our findings and developments pave the way toward a robust CL analyzer for exploring charge and energy transfer mechanisms in semiconductors, two-dimensional materials, and hybrid systems.

\section{Results and discussion}
\label{sec:resultsdiscussion}

\begin{figure}[htbp]
    \centering
    \includegraphics[width=\textwidth]{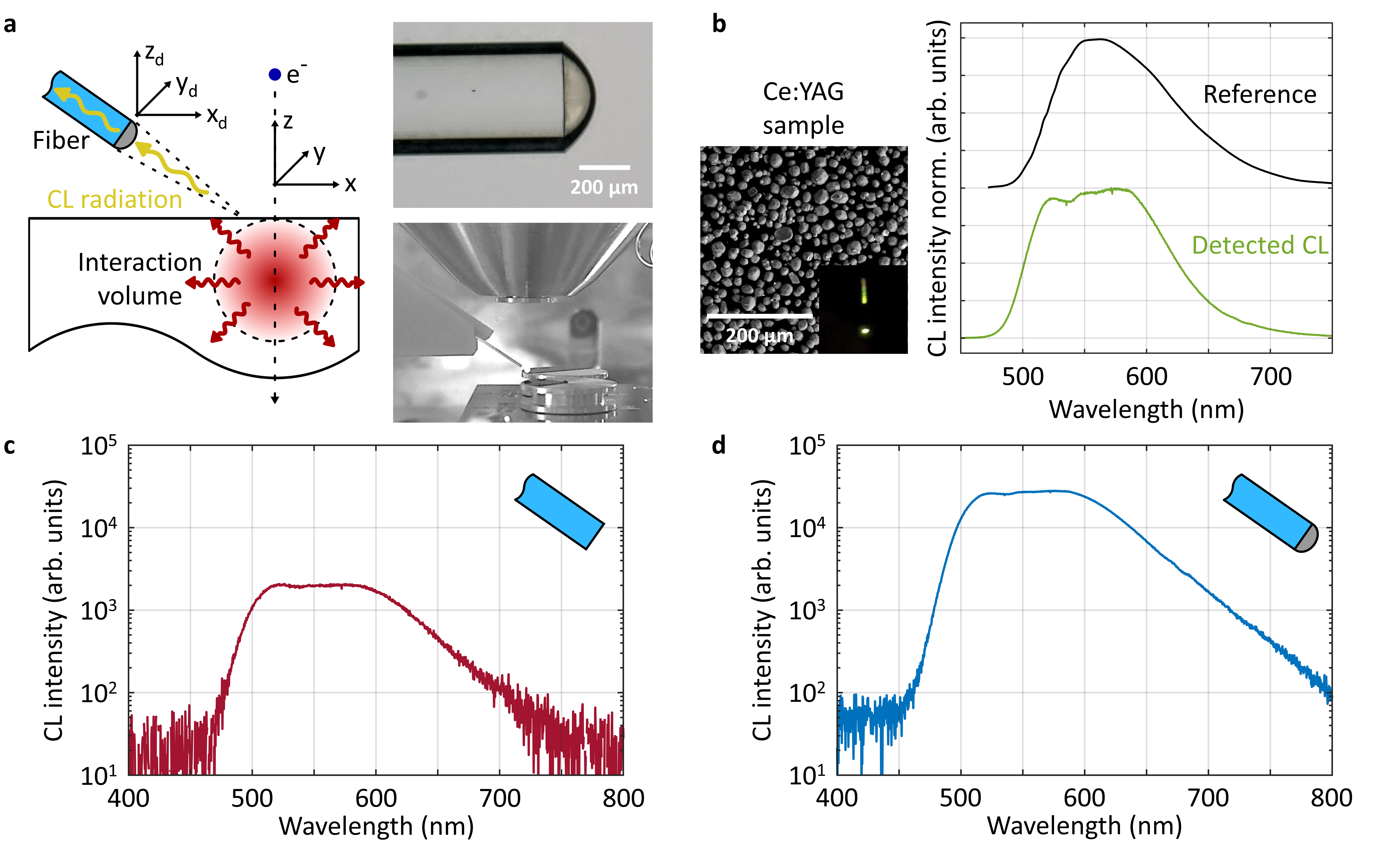}
    \caption{Schematic of the experimental setup for a fiber-based cathodoluminescence~(CL) detection. (a) An electron~($e^-$) with the kinetic energy of $eU_\text{acc}$ excites a variety of quasiparticles and collective excitations, e.g., excitons and surface plasmon polaritons, and generates CL emission. A multimode optical fiber with a core diameter of \SI{400}{\micro\meter} is connected to a three-axis piezo stage allowing for a precise positioning of the fiber collector with respect to the sample. The two insets on the right are showing an image of the uncoated multimode fiber and its positioning with respect to the sample as a side view. At the front face of the fiber, a lens is printed to improve the numerical aperture from $NA = \SI{0.22}{}$ up to $NA = \SI{0.36}{}$ for enhancing the collection efficiency of the emitted CL radiation. (b) Comparison of the measured CL spectrum of the Cerium-doped \ce{YAG} sample (\ce{Ce}:\ce{YAG}) with its referenced emission spectrum for $U_\text{acc} = \SI{30}{\kilo\volt}$. The insert on the left shows the secondary electron image of the particles and the visible bright luminescence of the \ce{Ce}:\ce{YAG} sample. (c) Measured CL spectrum taken with a flat cut multimode fiber. (d) Measured CL spectrum taken with a multimode fiber with 3D printed micro-optics, which has a circa ten times higher intensity count compared to the flat cut.}
    \label{fig:figure1}
\end{figure}

For the investigation of the material and its optical response, CL spectroscopy is performed by gathering the emitted radiation, in our case with a lensed optical multimode fiber (\Cref{fig:figure1}a), and directing it to an analyzing path. In our experimental setup, we implemented a fiber-based CL detector inside the sample chamber of a scanning electron microscope (SEM). The multimode fiber has a core diameter of \SI{400}{\micro\meter} and is fixed to a holder, that keeps the fiber at an inclination angle of \SI{35}{\degree} with respect to the sample stage, mounted onto a 3-axis piezo stage for precise positioning. Both the sample stage and the fiber can be positioned individually by using their respective stages, thereby our setup features noteworthy six degrees of freedom for movement. On top of this, the position of the electron beam and its spot size is further controlled via electron optics.

For a first functionality test, the CL spectrum of a Cerium-doped \ce{YAG} sample (\ce{Ce}:\ce{YAG}), that exhibits a bright luminescence, was measured with a cleaved multimode fiber and compared to the spectral shape of the fluorescence provided by the manufacturer (\Cref{fig:figure1}b) (Litec-LLL). The reference spectrum was obtained under broadband optical excitation, whereas our measurements are based on localized electron-beam excitation. This difference in excitation mechanism can lead to slight spectral variations due to local heterogeneities and the selective excitation of fewer emitters. Nevertheless, the overall agreement in spectral shape confirms the expected luminescence characteristics of the \ce{Ce}:\ce{YAG} sample. \ce{Ce}:\ce{YAG} has the molecular structure of \ce{Y3Al5O12} in the form of a cubic crystal and is provided in powder. The chemical stability of this material upon electron-beam irradiation and its strong luminescence made it a proper candidate for aligning and testing our CL apparatus. \ce{Ce}:\ce{YAG} with the bandgap of \SI{4.7}{\electronvolt} \cite{Xu1999} belongs to the class of large bandgap semiconductors with applications in optoelectronics and light-emitting diodes \cite{Chen2014}. The measured CL spectrum from this sample matches well with the reference, despite having a broader emission maximum around \SI{550}{\nano\meter}. The emission peak at \SI{550}{\nano\meter} is attributed to the $5d-4f$ emission of \ce{Ce^3+} ion \cite{Herrmann2015} and has been shown to be strongly influenced by the phase of the material, namely being in the crystalline or glass phases \cite{Han2019}. This slight difference between the measured CL spectrum and the provided fluorescence by the manufacturer is attributed to inhomogeneous broadening associated with the efficient excitation of different classes of defects in this material with electron beams \cite{Selim2007}.

To improve the collection efficiency of the fiber, a micro-optic lens was 3D printed on its apex, increasing the numerical aperture of the fiber from $NA=\SI{0.22}{}$ up to $NA=\SI{0.36}{}$. With the lens at the interface, the fiber now has a working distance (i.e., the distance from the lens apex to the focus) of approximately \SI{400}{\micro\meter} and a total acceptance angle of $\theta = \SI{42.2}{\degree}$, resulting in an improvement of ten times higher measured CL intensities for the same electron beam settings (\Cref{fig:figure1}c,d). For both measurements, the focal point of the fiber was fixed to the impact position of the electron beam, where the lateral position along the $x$ and $y$-axes was aligned by measuring the distance using the secondary electron detector of the electron microscope, and the vertical position was aligned by measuring the position at which the CL exhibits a maximum intensity.

\begin{figure}[htbp]
    \centering
    \includegraphics[width=\textwidth]{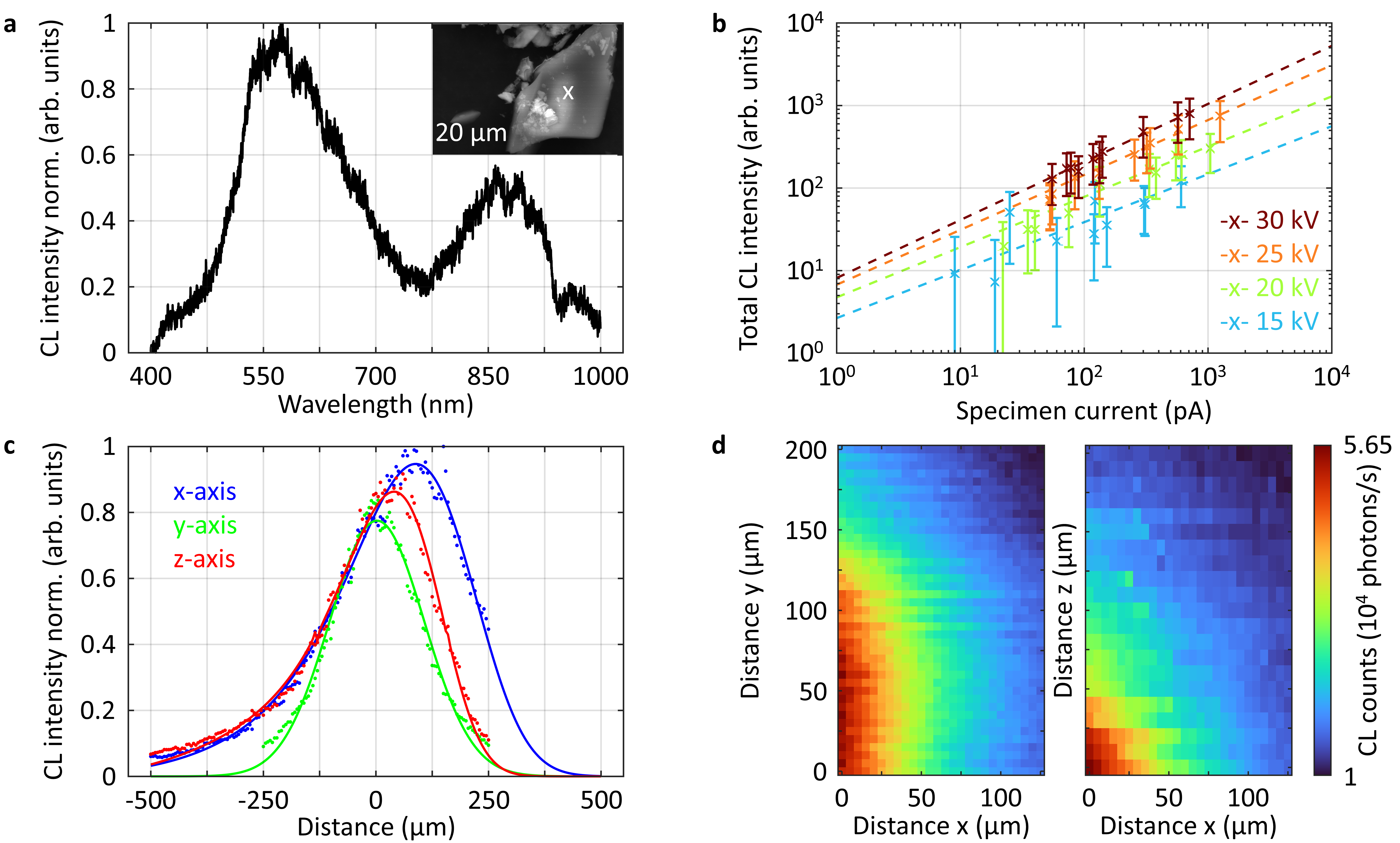}
    \caption{Cathodoluminescence~(CL) spectral measurements and the emission profile of a \ce{WSe2} flake. (a) CL intensity spectrum of the \ce{WSe2} flake acquired with the lensed fiber collector. The inset depicts a secondary electron image of the measured \ce{WSe2} flake, where the x marks the position of the electron beam excitation. The acceleration voltage of the electron beam was set to $U_\text{acc} = \SI{30}{\kilo\volt}$. The \ce{WSe2} flakes were exfoliated onto an adhesive carbon film. (b) The total CL intensity plotted versus the electron beam current $I_\text{beam}$ at the specimen location for different accelerations voltages $U_\text{acc}$ of the electrons. (c) One-dimensional CL position scans of the optical fiber along its three movement axes. The collected CL intensity decreases after the fiber is shifted out of the initial focus position (zero point). The intensity decrease shows a symmetric Gaussian shape in the $y$-direction, whereas an asymmetric behavior in $x$- and $z$-directions. (d) Two-dimensional CL mapping by moving the fiber in the $xy$-plane (left) or $xz$-plane (right) to illustrate the CL radiation profiles of \ce{WSe2}. The initial focus position of the fiber refers to the origin of coordinates.}
    \label{fig:figure2}
\end{figure}

After this initial setup and alignment of the fiber-based CL detector, the optical properties of the transition-metal dichalcogenide~(TMDC) semiconductor \ce{WSe2} was investigated, since it exhibits exciton excitations at room temperature \cite{Taleb2022,Selig2016,Lingstaedt2021,Wu2018,Wang2016,Munkhbat2019}. First, the CL spectrum of a \ce{WSe2} flake was measured, by collecting the radiation with the lensed fiber (\Cref{fig:figure2}a) and directing the radiation toward a grating-based spectrograph (Teledyne, Princeton Instrument) and projecting it onto a nitrogen-cooled charge-coupled device sensor. The spectrum features two dips at the wavelengths of approximately \SI{750}{\nano\meter} and \SI{860}{\nano\meter}, and a shallow peak at the wavelength of \SI{590}{\nano\meter}, where the latter is due to the excitation of B~excitons. The prominent dip at the wavelength of \SI{750}{\nano\meter} is associated with the A~exciton excitations of multilayer \ce{WSe2} films, well agreeing with the observations reported elsewhere \cite{Borghi2024,Woo2023,Ebel2024}. The second prominent spectral dip at the wavelength of \SI{860}{\nano\meter} is attributed to the excitation of momentum-dark excitons at the onset of $Q$ and $\Gamma$ transition \cite{Lindlau2018}. Notably, the large momentum transfer from the electron beams to the bound electrons has the potential to further excite and explore momentum-dark excitons in van der Waals materials and semiconductors. Instead of producing a luminescence peak, the exciton energies in \ce{WSe2} appear as dips in the CL spectra due to resonant absorption of broadband Cherenkov radiation and photonic modes by the excitons. This interaction can range from weak to strong coupling, forming exciton-polariton states, and is strongly dependent on both the flake thickness and the acceleration voltage of the electron beam \cite{Chahshouri2022,Taleb2022}. Overall, the excellent agreement between the CL measurements here and our previously acquired CL spectra using a commercial apparatus, highlights the accuracy of our fiber-based CL detector for characterizing the materials excitations.

Second, the electron beam parameters were altered to study their influence on the measured CL intensity by changing the electron beam current and the acceleration voltage of the electrons (\Cref{fig:figure2}b). For these measurements, the emitted photons were collected by the fiber and guided towards single-photon detectors (Hybrid Photomultiplier Detector Assembly, PicoQuant). The detector has a bandwidth of \SI{450}{\nano\meter} (total spectral range from \SIrange{450}{900}{\nano\meter}), gathering the entire spectral range of interest for exciton excitations in \ce{WSe2}. The measurement for each acceleration voltage indicates, that the total measured CL intensity~$I_\text{CL,total}$ follows a power function of the form
\begin{equation}
    I_\text{CL,total}(I_\text{beam}) = cI_\text{beam}^n
\end{equation}
with respect to the electron beam current~$I_\text{beam}$ at the specimen location. Here $c$ is a constant and the $n$ value depends on the acceleration voltage and ranges from $n=\SI{0.58}{}$ to $n=\SI{0.70}{}$ for the acceleration voltage changing from \SI{15}{\kilo\volt} up to \SI{30}{\kilo\volt}, respectively. The strong nonlinear dependence of the CL intensity on the electron beam current predominantly demonstrates the role of cascaded secondary interactions on the excited CL signal. The CL intensity as well demonstrates a strong dependence on the acceleration voltage. For higher acceleration voltages, the electron beam penetrates deeply inside the material and initiates even more cascaded inelastic interactions and therefore causes a higher number of emitted photons, as demonstrated also using Monte-Carlo simulations \cite{Chahshouri2022}.

Third, the optical multimode fiber can be moved along the three spatial directions with respect to the sample stage. To begin with, the CL intensity was measured while scanning the fiber along its three movement axes separately (\Cref{fig:figure2}c), where the collected CL intensity decreases while the fiber is shifted out of its initial focus position. Scanning along the $y$-direction, the CL intensity shows a symmetric Gaussian shape, whereas along the $x$- and $z$-direction there is an antisymmetric behavior visible, which is proven by fitting either a Gaussian or else a pseudo-Voigt function in form of
\begin{equation}
    f(r,a) = m \cdot \text{Gauss}(r \cdot p(r,a)) + (1-m) \cdot \text{Lorentz}(r \cdot p(r,a))
\end{equation}
to the respective data points. Here, $m$ is the contribution of the Gaussian function ($0 \leq m \leq 1$), $a$ is the asymmetry parameter and $p(r,a)$ is an additional perturbation term with a damped sigmoidal shape\cite{Korepanov2018}. All three fits exhibit values for $m=\SI{0.999}{}$ or higher and therefore indicating a CL intensity shape that is dominated by the Gaussian function, but with different asymmetry parameters for each one-dimensional scan. The fit for scanning along the $y$-direction has a relatively vanishing asymmetry parameter of $a=\SI{8.3e-4}{}$, whereas $a=\SI{-0.33}{}$ and $a=\SI{-0.39}{}$ for the $x$- and $z$-direction, respectively. The sign of the asymmetry parameter refers to the inclination toward the right or left side of the movement axes.

The main interest is to scan the optical fiber along multiple dimensions to reveal the CL radiation profiles. Notably, the fiber-based CL detector collects far-field radiation emitted from the electron-beam excitation spot, and by raster-scanning the fiber, it records the spatial distribution of the radiation pattern. In the case of semiconductors with sub-micrometer exciton diffusion lengths, the CL maps effectively represent the angular emission profile of a point-like source. This enables direct imaging of the radiation pattern in Fourier space, rather than a real-space image or diffusion map. Moreover, by scanning the fiber laterally at a fixed height, the system can also detect secondary emission centers, providing insights into exciton diffusion and energy transport in complex heterostructures \cite{Darbari2025}. For \ce{WSe2} the CL radiation profiles in the $xy$-plane and $xz$-plane are illustrated (\Cref{fig:figure2}d). The profile in the $xy$-plane displays an elliptical shape, whereas the $xz$-plane displays a spherical shape which can be connected to the expected pseudo-Lambertian radiation profile of a semiconductor like \ce{WSe2}. The radiation profile deviates from a pure Lambertian form due to the coexistence of coherent and incoherent radiation mechanisms \cite{Brenny2014}.

\begin{figure}[htbp]
    \centering
    \includegraphics[width=\textwidth]{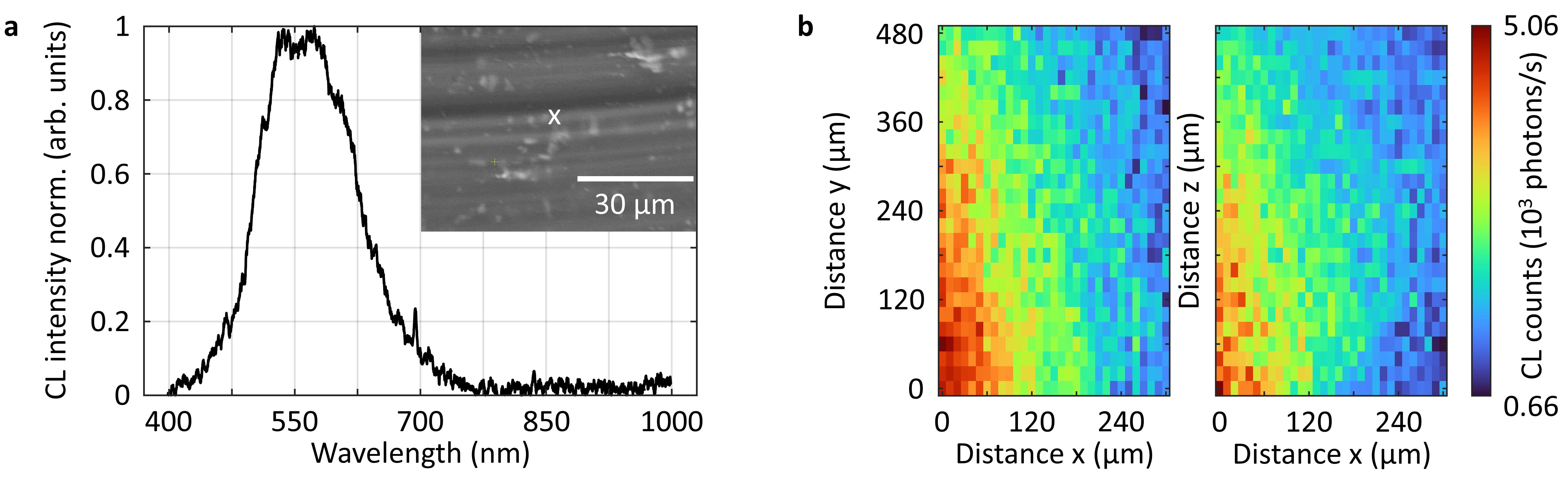}
    \caption{Cathodoluminescence~(CL) spectral measurements and emission profile imaging of a gold thin film with a thickness of \SI{40}{\nano\meter}, deposited on a stub. (a) CL intensity spectrum of the gold sample. The insert depicts a SEM image of the gold sample, where the x marks the position of the electron beam excitation. The acceleration voltage of the electron beam was set to $U_\text{acc} = \SI{30}{\kilo\volt}$. (b) Two-dimensional CL mapping by moving the fiber in the $xy$-plane (left) or $xz$-plane (right) to illustrate the CL radiation profiles of gold. The initial focus position of the fiber refers to the origin of coordinates.}
    \label{fig:figure3}
\end{figure}

To explore the radiation profile of transition radiation, another sample consisting of a thin \SI{40}{\nano\meter} gold layer was prepared. Again, the CL spectrum was measured, indicating a broad emission inside the visible spectrum from \SIrange{500}{650}{\nano\meter} (\Cref{fig:figure3}a), attributed to the transition radiation \cite{Ebel2024}. For a comparison between the semiconductor \ce{WSe2} and the metallic gold layer, the CL radiation profile was measured by scanning the fiber in the two-dimensional $xy$-plane and $xz$-plane (\Cref{fig:figure3}b). In the $xy$-plane the CL radiation profile of the metallic sample looks similar compared to the elliptical shape of the semiconductor. But in case of the $xz$-plane scans, the CL radiation profile changes and now looks like a cut through the expected toroidal shape.

\begin{figure}[htbp]
    \centering
    \includegraphics[width=\textwidth]{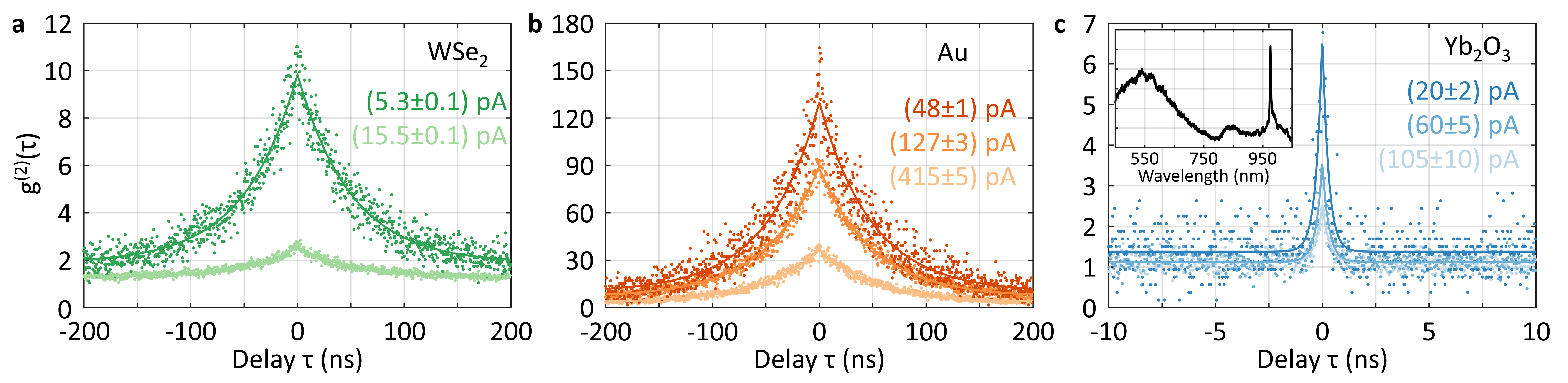}
    \caption{Second-order autocorrelation function $g_\text{CL}^{(2)}(\tau)$ measurements of the CL intensity~$I_\text{CL}(t)$ for the \ce{WSe2} flake (a), a thin evaporated gold layer (b), and Ytterbium~(III) oxide (\ce{Yb2O3}) (c). In each measurement, the acceleration voltage of the electron beam was fixed at $U_\text{acc} = \SI{30}{\kilo\volt}$, whereas the beam current $I_\text{beam}$ at the specimen location was altered to demonstrate the enhancement of the bunching amplitude at lower currents. The insert illustrates the measured CL spectrum of the \ce{Yb2O3} sample and its narrow resonance in the infrared range around \SI{975}{\nano\meter}.}
    \label{fig:figure4}
\end{figure}

For further investigating the decay time of the excitation explored here, a fiber-based HBT intensity interferometer was constructed, which consists of a multimode fiber-based beam splitter and two photomultiplier tubes, connected to an electronic correlator (quTAG standard time tagger with \SI{10}{\pico\second} timing jitter). Using this setup, we perform time-correlated single photon counting and measure the second-order autocorrelation function~$g_\text{CL}^{(2)}(\tau)$ of the measured CL intensity signal~$I_\text{CL}(t)$ \cite{Scheucher2022,Fiedler2023a,Feldman2018,Meuret2018}. This gives an insight into the interaction of high-energy electrons with matter and the coherent and incoherent processes behind them. Each electron excites multiple quantum states in an extremely short interaction time of less than \SI{1}{\femto\second}. This short interaction time establishes an inherent time synchronization between the emitter, hence resulting in an observed bunching effect. The excitations diffuse and recombine radiative over time \cite{Meuret2015}, and therefore, the second-order autocorrelation function features a photon bunching effect at zero delay ($g_\text{CL}^{(2)}(0) \gg 1$) in CL experiments, within the available temporal resolution of the HBT setup explored here. The radiative lifetime of the ensembles of the emitters can be determined by the decay rate of the bunching peak. Here, the second-order autocorrelation functions of the previously explored materials, namely a \ce{WSe2} flake and a thin gold layer, and Ytterbium oxide (\ce{Yb2O3}), were measured (\Cref{fig:figure4}). The latter one has a narrow resonance in the infrared range around \SI{975}{\nano\meter}. During the measurement the exciting electrons have a fixed acceleration voltage of \SI{30}{\kilo\volt}, whereas the beam current $I_\text{beam}$ at the specimen location was altered to demonstrate the enhancement of the bunching amplitude at lower currents \cite{Fiedler2023b}. For each sample and beam current, the model function
\begin{equation}
    g_\text{CL}^{(2)}(\tau) = 1 + g_0 \cdot \exp( -\vert\tau\vert / \tau_\text{d} )
\end{equation}
was fitted to the experimental data, where $g_0$ represents the amplitude of $g_\text{CL}^{(2)}$ at the delay of $\tau=0$ and $\tau_\text{d}$ is the decay time of the excited state, or its lifetime respectively. The extracted lifetimes associated with each measurement are shown at \Cref{tab:table1}. In all the explored materials, the bunching peak is enhanced for lower electron beam currents. Surprisingly, the emission from the gold film exhibits the largest bunching peak, even larger than semiconductors and a luminescence material like \ce{Yb2O3}. This effect demonstrates the role of SPP and perhaps bulk plasmons in further synchronizing the emissions over a temporally short time scale, associate with the decay time of bulk plasmons. Although the intrinsic relaxation times of transition radiation and SPP are in the femtosecond regime and beyond the temporal resolution of our $g_\text{CL}^{(2)}$ measurements, the electron beam also excites longer-lived interband and intraband transitions in the gold film, resulting in an ensemble of secondary emitters that dominate the measured CL signal at longer timescales.

\begin{table}[]
    \centering
    \begin{tabular}{c|c|c|c|c|c|c|c|c}
        Material & \multicolumn{2}{c|}{\ce{WSe2}} & \multicolumn{3}{c|}{\ce{Au}} & \multicolumn{3}{c}{\ce{Yb2O3}} \\\hline 
        Current~$I_\text{beam}~(\si{\pico\ampere})$ & 5.3 & 15.5 & 48 & 127 & 415 & 20 & 60 & 105\\
        Deacy time~$\tau_\text{d}~(\si{\nano\second})$ & 49.8 & 49.1 & 49.2 & 50.9 & 53.6 & 0.28 & 0.29 & 0.32\\
        Bunching peak $g_0$ & 8.0 & 1.3 & 120 & 81.9 & 33.5 & 5.4 & 2.5 & 1.4
    \end{tabular}
    \caption{Decay time and bunching peak of the electron-induced optical emission process in different materials.}
    \label{tab:table1}
\end{table}

\section{Conclusion}
\label{sec:conclusion}
In conclusion we have shown first proof-of-concept experiments, that demonstrate the functionality and flexibility of our developed fiber-based cathodoluminescence detector. Its versatility is highlighted by resolving the far-field spatial distribution of the emitted light, as well as performing cathodoluminescence spectroscopy and time-correlated single photon counting. Moreover, we could increase the light collection efficiency of the optical multimode fiber by 3D printing a micro-optic lens on its front side. Using this CL detector, certain emission sites on the sample can be selectively located and their diffusion properties be tracked over space.

\section{Acknowledgement}
\label{sec:acknowledgement}

We gratefully acknowledge collaborations with the SmarAct company to design the nano positioner system holding the fiber and Toon Coenen (AMOLF, Amsterdam) for fruitful discussions for realizing a fiber-based CL system and with Sven Ebel (Southern Denmark University) for discussions regarding CL measurements. We acknowledge Kai Rossnagel (Kiel University, Germany) for providing us with \ce{WSe2} samples. This project has received funding from the Volkswagen Foundation (Momentum Grant), European Research Council~(ERC) under the European Union’s Horizon 2020 research and innovation program under grant agreement no.~101157312 (UltraCoherentCL), grant agreement no.~101017720 (EBEAM), grant agreement no.~101170341 (UltraSpecT), and from Deutsche Forschungsgemeinschaft. P.R. and H.G. acknowledge support from Bundesministerium für Bildung und Forschung (3DPrintedOptics, Integrated3Dprint, QR.X, QR.N) and Deutsche Forschungsgemeinschaft (DFG, German Research Foundation, 431314977/GRK2642).

\section{Methods}
\label{sec:methods}

\subsection{Optical fiber and 3D printed micro-optics}
\label{ssec:opticalfiber}

For our experiments we were using the optical multimode fiber FG400AEA from Thorlabs with a silica core diameter of \SI{400\pm8}{\micro\meter} and a broad spectral operation range from \SIrange{180}{1200}{\nano\meter}. The pristine fiber with a flat cross section has a numerical aperture of \SI{0.22\pm0.02}{}. For improving the numerical aperture, and therefore the collection efficiency, a lens was 3D printed onto the face of the multimode fiber. We employed a two-photon polymerization 3D printing \cite{Gissibl2016,Ruchka2022} utilizing machine Nanoscribe QuantumX and a photopolymer IP-S (Nanoscribe GmbH). We used the grayscale printing mode with a $25x$~objective, which provides especially smooth surfaces and thus enhances the performance of the micro-optics \cite{Siegle2023}. After printing, the fiber with the lens was developed for \SI{15}{\minute} in mr-Dev600 (micro resist technology GmbH) and rinsed for \SI{3}{\minute} in IPA. We also UV-cured the optics \cite{Schmid2019} to ensure complete polymerization and hence the homogeneity of the optical properties of the lens. With this micro-optic the numerical aperture of the fiber was increased to \SI{0.36}{} and resulting in a working distance of ca. \SI{400}{\micro\meter}. On the downside, the additional printed lens decreased the transmission efficiency in the spectral range of \SI{400}{\nano\meter} up to \SI{450}{\nano\meter} to less than \SI{30}{\percent}, but across the visible spectral range above the transmission efficiency remained at approximately \SI{95}{\percent}.

\subsection{Cathodoluminescence spectroscopy}
\label{ssec:CLspectroscopy}

All the measurements shown here were performed using the Thermo Fisher Quattro S field-emission scanning electron microscope~(FE-SEM). Throughout the experiments, the acceleration voltage and the current of the electron beam was altered as specified in the main text. The sample stage of the SEM is connected to an in-built amperemeter to measure the effective beam current at the specimen location. On the inside of the SEM, the optical multimode fiber was mounted on a system of three linear piezo stages from SmarAct, which feature an accuracy of \SI{1}{\nano\meter} in step size. Hence, the fiber is movable independently from the sample stage along three axial degrees of freedom, including a total dynamic range of \SI{83}{\milli\meter} along the $x$-axis and \SI{35}{\milli\meter} along $y$- and $z$-axis, for a precise spatial collection of the cathodoluminescence~(CL) radiation. Thanks to its flexibility, the fiber cable will guide the collected CL radiation towards a spectrometer or photomultiplier tubes, respectively. Initially we used the Avantes AvaSpec-ULS4096CL-EVO spectrometer for measuring the luminescent \ce{Ce}:\ce{YAG} sample, which provides a wavelength range of \SIrange{200}{1100}{\nano\meter}, groove density of \SI{300}{\per\milli\meter}, blaze wavelength of \SI{300}{\nano\meter} and a resolution of \SI{0.6\pm0.1}{\nano\meter}. Afterwards, the setup was upgraded with the Teledyne Princeton Instruments HRS-500 spectrograph attached with the PyLoN~100BRX liquid-nitrogen-cooled CCD camera. This spectrograph is equipped with three gratings, which all have a groove density of \SI{1200}{\per\milli\meter} with a spectral resolution of \SI{0.1}{\nano\meter} and blaze wavelengths of \SI{300}{\nano\meter}, \SI{500}{\nano\meter}, and \SI{750}{\nano\meter}, respectively.

\subsection{Time-correlated single photon counting}
\label{ssec:TCSPC}

Additionally, an assembly consisting of two PMA Hybrid~50 photomultiplier tubes from PicoQuant and the quTAG time-to-digital converter from qutools is built up for measuring the CL count rate. By using the TT400R5F1B 1x2 multimode fiber coupler from Thorlabs with a 50:50 splitting ratio, this assembly operates as a Hanbury Brown-Twiss intensity interferometer for performing time-correlated single photon counting and measuring the second-order autocorrelation function $g_\text{CL}^{(2)}(\tau)$ of the emitted CL intensity signal.

\subsection{Sample preparation}
\label{ssec:samplepreparation}

The Cerium-doped \ce{YAG} sample (\ce{Ce}:\ce{YAG}) was purchased from Litec-LLL and put onto an adhesive carbon tape. The \ce{WSe2} flakes were grown via the standard chemical vapor transport method and then mechanically exfoliated onto an adhesive carbon tape. The gold sample was made by the thermal evaporation of gold onto a commercial SEM stub in a PVD~chamber. Here, the evaporated gold layer has a thickness of roughly a few tens of nanometers. Lastly, a shard of the Ytterbium~(III) oxide (\ce{Yb2O3}) sample was glued on carbon tape.
\clearpage

\clearpage

\bibliography{Manuscript_Fiber_CL}
\clearpage

\end{document}